\documentclass[11pt,twoside]{article}
\usepackage[a4paper]{geometry}
\usepackage{verbatim}
\usepackage{float}
\usepackage{amsmath}
\usepackage{amssymb}
\usepackage{graphicx}
\usepackage{multirow}
\usepackage{here}

\newcommand {\beq} {\begin{equation}}
\newcommand {\eeq} {\end{equation}}
\newcommand {\beqa}{\begin{eqnarray}}
\newcommand {\eeqa}{\end{eqnarray}}
\newcommand {\nn} {\nonumber}

\newcommand {\tr}{{\rm tr\,}}

\newcommand {\ee}{\mbox{e}}

\newcommand{\bbC}{{\mathbb C}}

\makeatletter

\numberwithin{equation}{section}

\usepackage[verbose]{cite}

\makeatother

\date{}
\begin{document}

\begin{flushright} 
KEK-TH-1934
\end{flushright} 

\vspace{1cm}

\begin{center}
{\LARGE The complex Langevin analysis of spontaneous symmetry
breaking induced by complex fermion determinant}
\end{center}
\vspace{0.1cm}
\vspace{0.1cm}
\begin{center}

\vspace{1cm}

         Yuta I{\sc to}$^{a}$\footnote
          {
 E-mail address : yito@post.kek.jp}
and
         Jun N{\sc ishimura}$^{ab}$\footnote
          {
 E-mail address : jnishi@post.kek.jp} 

\vspace{1.5cm}

$^a${\it KEK Theory Center, 
High Energy Accelerator Research Organization,\\
1-1 Oho, Tsukuba, Ibaraki 305-0801, Japan}

$^b${\it Graduate University for Advanced Studies (SOKENDAI),\\
1-1 Oho, Tsukuba, Ibaraki 305-0801, Japan} 

\end{center}

\vspace{1.5cm}

\begin{center}
  {\bf abstract}
\end{center}

\noindent In many interesting physical systems, the determinant
which appears from integrating out fermions becomes complex,
and 
its phase
plays a crucial role in the determination of the vacuum. 
An example of this is
QCD at low temperature and high density, where various
exotic fermion condensates are conjectured to form.
Another example is
the Euclidean version of the type IIB matrix model for
10d superstring theory, where
spontaneous breaking of the SO(10) rotational symmetry 
down to SO(4) is expected to occur.
When one applies the complex Langevin method to these systems,
one encounters the singular-drift problem associated with 
the appearance of nearly zero eigenvalues of the Dirac operator.
Here we propose to avoid this problem by deforming the action
with a fermion bilinear term.
The results for the original system are obtained 
by extrapolations with respect to the deformation parameter.
%
We demonstrate the power of this approach
by applying it to a simple matrix model, in which
spontaneous symmetry breaking from SO(4) to SO(2) is expected 
to occur due to the phase of the complex fermion determinant.
Unlike previous work based on a reweighting-type method,
we are able to determine the true vacuum by
calculating the order parameters,
which agree with the prediction by the Gaussian expansion method.


\newpage

\section{Introduction\label{sec:Introduction}}

The sign problem is a notorious technical problem that occurs
in applying Monte Carlo methods to a system with a complex action $S$.
The importance sampling cannot be applied as it is
since the integrand $\exp\left(-S\right)$ of the partition
function cannot be regarded as a Boltzmann weight.
If one uses the absolute value $|\exp\left(-S\right)|$ for generating
configurations and treats the phase factor as a part of the observable,
huge cancellations occur among configurations, and the required
statistics grows exponentially with the system size.
This problem occurs in various interesting systems in particle physics
such as finite density QCD, gauge theories with a theta term or 
a Chern-Simons term, chiral gauge theories and supersymmetric theories.

The complex Langevin method (CLM) \cite{Parisi:1984cs,Klauder:1983sp}
is a promising approach to such complex-action systems, which may be
regarded as an extension of the stochastic quantization 
based on the Langevin equation.
The dynamical variables of the original system 
are naturally complexified, and the observables
as well as the drift term are extended holomorphically by analytic continuation.
It is known that the CLM 
works beautifully in highly nontrivial cases
\cite{Aarts:2008rr,Aarts:2008wh,Aarts:2010gr,Aarts:2011zn},
while it gives simply wrong results in the other cases 
\cite{Ambjorn:1985iw,Ambjorn:1986fz,Aarts:2010aq,Pawlowski:2013pje}.

In the past several years, significant progress has been made
in theoretical understanding of 
the method
and the conditions for justifying the CLM.
First it was realized that the probability distribution of 
the complexified dynamical variables has to fall off fast enough 
in the imaginary directions of the configuration 
space \cite{Aarts:2009uq,Aarts:2011ax}.
In order to satisfy this condition,
a new technique called gauge cooling \cite{Seiler:2012wz}
was proposed.
Using the gauge cooling, the CLM has been successfully applied to
finite density QCD\footnote{There are also attempts 
to apply the CLM to the real-time dynamics
\cite{Berges:2005yt,Berges:2006xc,Berges:2007nr,Anzaki:2014hba} and
to Yang-Mills theory 
with a theta term \cite{Bongiovanni:2013nxa,Bongiovanni:2014rna}.}
either with heavy quarks \cite{Seiler:2012wz}
or at high temperature \cite{Sexty:2013ica}.
An explicit justification of the gauge cooling has been 
provided recently \cite{Nagata:2015uga} extending the 
argument for justification of the CLM 
without gauge cooling \cite{Aarts:2009uq,Aarts:2011ax}.

It was known for some time that 
the CLM gives wrong results also when the determinant that
appears from integrating out fermions
takes values close to zero during the complex Langevin simulation.
This was first realized
in the Random Matrix Theory for finite density QCD 
\cite{Mollgaard:2013qra,Mollgaard:2014mga}
and confirmed also in 
effective Polyakov line models \cite{Greensite:2014cxa}.
In these papers, it was speculated that the problem occurs 
due to the ambiguity associated with the branch cut in
the logarithm of the complex fermion determinant, 
which appears in the effective action.
On the other hand, ref.~\cite{Seiler}
pointed out that the singular drift term one obtains
from the fermion determinant breaks holomorphy, which plays a crucial role
in justifying the method.

A theoretical understanding of this problem and a possible cure 
have been given recently.
First it was pointed out in ref.~\cite{Nishimura:2015pba}
that the branch cut cannot be the cause of the
problem since the CLM can be formulated solely in terms of the weight $w=\exp(-S)$
without ever having to refer to the action $S$.
Indeed it was found that a similar problem can occur 
when the action has pole singularities instead of logarithmic singularities.
In the same paper, it was shown that
the probability distribution of the complexified variables
has to fall off fast enough near the singularities of
the drift term, based on the argument for justification
in ref.~\cite{Aarts:2009uq,Aarts:2011ax}.
It was then proposed \cite{Nagata:2015ijn,Nagata:2016alq} that 
the gauge cooling can be used to satisfy
this condition as well 
with an appropriate choice of the complexified gauge transformation. 
A test in the Random Matrix Theory shows that the gauge cooling indeed 
solves the singular-drift problem
unless the quark mass becomes too small.

In ref.~\cite{Nagata:2016vkn}, the argument for justification
with or without gauge cooling was revisited.
In particular, it was pointed out
that the expectation values of time-evolved observables,
which play a crucial role in the argument, can be ill-defined.
Taking this into account, 
it was shown that
the CLM can be justified
if
the probability distribution
of the drift term falls off exponentially or faster at large magnitude.
This condition serves as a useful criterion, which tells us
clearly whether the results obtained by the CLM are trustable or not.

In this paper, we focus on the singular-drift problem
that occurs in a system with a complex fermion determinant.
In many such systems,
the phase of the fermion determinant is expected to
play a crucial role in the determination of the vacuum.
%
An example of this is
finite density QCD at low temperature and high density,
where
various exotic fermion condensates are conjectured to
form (See ref.~\cite{Rajagopal:2000wf}, for instance.).
Another example is
the Euclidean version of 
the type IIB matrix model \cite{Ishibashi:1996xs}
for 10d superstring theory,
where 
the SO(10) rotational symmetry is conjectured to be spontaneously 
broken \cite{Aoki:1998vn,Nishimura:2000ds,%
Nishimura:2000wf,Anagnostopoulos:2001yb}.
When one applies the CLM to these systems,
the singular-drift problem occurs
due to the appearance of eigenvalues of the Dirac operator close to zero.
We propose to avoid this problem by deforming the action
with a fermion bilinear term and extrapolating its coefficient to zero.
The fermion bilinear term should be chosen in such a way that 
the nearly zero eigenvalues of the Dirac operator are avoided 
and yet
the vacuum of the system is minimally affected.

We test this idea in an SO(4)-symmetric matrix model
with a Gaussian action and a complex fermion 
determinant, in which spontaneous breaking of SO(4) symmetry
is expected to occur due to the phase of the 
determinant \cite{Nishimura:2001sq}.
This model was studied previously
by the Gaussian expansion method (GEM) \cite{Nishimura:2004ts}
and the spontaneous breaking of the SO(4) symmetry down to SO(2)
was suggested by comparing the free energy for the SO(2)-symmetric vacuum
and the SO(3)-symmetric vacuum.
The same model was studied also
by Monte Carlo simulation using the factorization 
method\footnote{This is a kind of reweighting method that attempts
to solve the so-called overlap problem, which is an important part of the 
complex-action problem. 
While the original version was proposed in ref.~\cite{Anagnostopoulos:2001yb},
the importance of constraining observables which
are strongly correlated with the phase of the determinant
was recognized later in 
refs.~\cite{Anagnostopoulos:2010ux,Anagnostopoulos:2011cn}.}, 
and 
the order parameters
obtained by the GEM
were reproduced 
for both the SO(2)-symmetric vacuum and the SO(3)-symmetric 
vacuum \cite{Anagnostopoulos:2010ux,Anagnostopoulos:2011cn}.
However, the comparison of free energy for the two vacua
suffered from too much uncertainty to make
a definite conclusion on the true vacuum by this approach.
%

When one applies the CLM to this system,
the singular-drift problem
is actually severe because the fermionic part of the model is 
essentially an exactly ``massless'' system.
Indeed, it turns out that the gauge cooling proposed 
in refs.~\cite{Nagata:2015ijn,Nagata:2016alq} 
is not sufficient to
solve this problem in the case at hand.
Following the idea described above,
we therefore add a fermion bilinear term, which breaks the SO(4) symmetry
minimally, down to SO(3).
The results of the CLM show 
that the SO(3) symmetry of the deformed model is
broken spontaneously to SO(2).
Extrapolating the deformation parameter to zero,
we find that
the SO(4) symmetry of the original matrix model
is broken spontaneously to SO(2)
and that the order parameters thus obtained
agree well with the prediction obtained by the GEM.
%
We also try another type of the fermion bilinear term
for the deformation and show that 
the final results obtained after the extrapolations remain the same,
which supports the validity of our analysis.
Note that
we are able to determine the true vacuum directly 
without having to compare the free energy for each vacuum preserving 
different amount of rotational symmetry.

In order to probe the spontaneous symmetry breaking (SSB),
we need to introduce an O($\varepsilon$) symmetry breaking term in the action,
on top of the deformation described above,
and send $\varepsilon$ to zero after taking the large-$N$ limit.
The singular-drift problem occurs
at small $\varepsilon$ even for the deformed model.
Here, the criterion for correct convergence
proposed recently \cite{Nagata:2016vkn} turns out to be useful
since it tells us which data are free from the singular-drift problem
and hence can be trusted.
Indeed, we find that the data points in the reliable region
can be fitted nicely by an expected asymptotic behavior, 
while the data points in the unreliable region
deviate from the fitting curve.
%
We hope that our strategy to overcome the singular-drift problem
enables 
the application of the CLM to
the type IIB matrix model
and to finite density QCD at low temperature and high density.

The rest of this paper is organized as follows. 
In section \ref{sec:The-definition-of},
we define the SO(4)-symmetric matrix model
and briefly review the results obtained by the previous approaches.
%
In section \ref{sec:Application-of-the}, we explain
how we apply the CLM to the SO(4)-symmetric matrix model.
In particular, we deform the action with a fermion bilinear term,
which enables us to investigate the SSB without suffering from
the singular-drift problem.
In section \ref{sec:results},
we present the results of our analysis.
In particular, we extrapolate the deformation parameter to zero,
and confirm that the SSB from SO(4) to SO(2) indeed occurs
in this model. The order parameters thus obtained are
in good agreement with the prediction of the GEM.
Section \ref{sec:Summary-and-discussion}
is devoted to a summary and discussions. 
In appendix \ref{sec:region-of-validity}
we give the details on how we determine the region
of validity of the CLM, which is useful in making
the $\varepsilon \rightarrow 0$ extrapolations.
In appendix \ref{sec:3rd-fermion-case}, 
we present the results obtained by deforming the action
with another type of the fermion bilinear term,
which turn out to be consistent
with the ones obtained in section \ref{sec:results}.

\section{Brief review of the SO(4)-symmetric matrix model
\label{sec:The-definition-of}}

The SO(4)-symmetric matrix model investigated in this paper
is defined by the partition function \cite{Nishimura:2001sq}
\begin{equation}
Z=\int dX d\psi d\bar{\psi} \, e^{-\left(S_{{\rm b}}+S_{{\rm f}}\right)},
\label{eq:partition_func}
\end{equation}
where the bosonic part and the fermionic part of the action is given, 
respectively, as
\begin{eqnarray}
S_{{\rm b}} & = & \frac{1}{2}N\sum_{\mu=1}^{4}{\rm tr} \, (X_{\mu})^{2} \ ,
\label{eq:boson_action}\\
S_{{\rm f}} & = & -N \sum_{f=1}^{N_{\rm f}}
\sum_{\mu=1}^{4}\sum_{\alpha,\beta=1}^{2} 
\bar{\psi}_{\alpha}^{(f)}
\left(\Gamma_{\mu}\right)_{\alpha\beta}X_{\mu} \, 
\psi_{\beta}^{(f)} 
 \ .
\label{eq:fermion_action}
\end{eqnarray}
Here we have introduced 
$N\times N$ Hermitian matrices $X_{\mu}$ $(\mu=1,\ldots,4)$, which are bosonic,
and $N_{\rm f}$ copies of
$N$-dimensional column vectors $\psi_{\alpha}^{(f)}$
and 
row vectors $\bar{\psi}_{\alpha}^{(f)}$
$(f=1 , \ldots , N_{\rm f}; \  \alpha=1,2)$, 
which are fermionic.
The $2\times2$ matrices $\Gamma_{\mu}$ are the
gamma matrices in 4d Euclidean space after Weyl projection, 
which are defined by 
\[
\Gamma_{\mu}=\begin{cases}
i \, \sigma_{i} & \text{for}\; \mu=i=1 , 2, 3 \ ,\\
\mathbf{1}_{2} & \text{for~}\mu=4 \ ,
\end{cases}
\]
using the Pauli matrices $\sigma_{i}$ ($i=1,2,3$).
The model has an SO(4) symmetry,
under which $X_{\mu}$ transforms as a vector, whereas
$\psi_{\alpha}$ and $\bar{\psi}_{\alpha}$ transform as Weyl spinors. 
Also, the model has an SU($N$) symmetry, under which the dynamical variables
transform as
\begin{alignat}{3}
X_\mu & \mapsto g \, X_\mu \, g^{-1}
\ , \quad\quad
\psi_{\alpha}^{(f)} & \mapsto g \, \psi_{\alpha}^{(f)} 
\ , \quad\quad
\bar{\psi}_{\alpha}^{(f)} & \mapsto \bar{\psi}_{\alpha}^{(f)} g^{-1} \ ,
\end{alignat}
where $g \in {\rm SU}(N)$.

Integrating out the fermionic variables for each $f$, 
one obtains the determinant of the Dirac operator
\begin{eqnarray}
D_{i\alpha , j \beta} & = & \sum_{\mu=1}^{4} (\Gamma_{\mu})_{\alpha\beta} 
(X_{\mu})_{ij} \ ,
\label{eq:dirac_op}
\end{eqnarray}
which is complex in general.
Thus, the partition function (\ref{eq:partition_func}) can be rewritten as
\begin{equation}
Z=\int dX \,\left(\det D\right)^{N_{\rm f}}e^{-S_{{\rm b}}} \ .
\label{part-fn-with-det}
\end{equation}
It was speculated that the SO(4) rotational symmetry of the model is
spontaneously broken in the large-$N$ limit with fixed $r=N_{\rm f}/N >0$
due to the effect of the phase of the determinant \cite{Nishimura:2001sq}.
In the phase-quenched model, which is defined by omitting 
the phase of the fermion determinant,
the SSB was shown not to occur
by Monte Carlo simulation \cite{Anagnostopoulos:2011cn}.
We may therefore say that the SSB, if it really occurs,
should be induced by the phase of the fermion determinant.
Throughout this paper, we consider the $r=1$ case, which corresponds
to $N_{\rm f}=N$.

In order to see the SSB,
we introduce an SO(4)-breaking mass term 
\begin{equation}
\Delta S_{{\rm b}}= \frac{N}{2} \varepsilon \sum_{\mu=1}^{4}
 m_{\mu} {\rm tr} \, (X_{\mu})^{2} 
\label{eq:boson_action_bdeform}
\end{equation}
in the action, where
\begin{equation}
m_1 < m_2 < m_3 < m_4  \ ,
\label{eq:boson_mass_split}
\end{equation}
and define the order parameters for the SSB
by the expectation values of
\begin{alignat}{3}
\lambda_\mu
=\frac{1}{N}{\rm tr} \,  (X_{\mu})^2  \ ,
\label{lambda-def}
\end{alignat}
where no sum over $\mu$ is taken.
Due to the ordering (\ref{eq:boson_mass_split}),
the expectation values obey
\begin{alignat}{3}
\langle \lambda_1 \rangle > 
\langle \lambda_2 \rangle > 
\langle \lambda_3 \rangle > 
\langle \lambda_4 \rangle 
\label{lambda-ordering}
\end{alignat}
at finite $\varepsilon$.
Taking the large-$N$ limit
and then sending $\varepsilon$ to zero afterwards,
the expectation values $\langle \lambda_\mu \rangle$ ($\mu=1,\cdots , 4$)
may not take the same value. 
In that case, we can conclude that the SSB occurs.

Explicit calculations based on the 
GEM
were carried out assuming that the SO(4) symmetry is broken down 
either to SO(2) or to SO(3) \cite{Nishimura:2004ts}. 
For $r=1$,
the order parameters are given by
\begin{alignat}{3}
 \left\langle \lambda_{1}\right\rangle
=\left\langle \lambda_{2}\right\rangle  \sim 2.1 \ ,
\quad \left\langle \lambda_{3}\right\rangle  \sim 1.0 \ ,
\quad \left\langle \lambda_{4}\right\rangle  \sim 0.8  
\quad \quad  & \mbox{for the ${\rm SO(2)}$ vacuum} \ ,
\label{eq:previous_result}
\\
\left\langle \lambda_{1}\right\rangle
=\left\langle \lambda_{2}\right\rangle
= \left\langle \lambda_{3}\right\rangle  \sim 1.75 \ ,
\quad \left\langle \lambda_{4}\right\rangle  \sim 0.75  
\quad\quad   & \mbox{for the ${\rm SO(3)}$ vacuum} \ .
\label{eq:previous_result_SO3}
\end{alignat}
The free energy was calculated in each vacuum,
and the SO(2)-symmetric vacuum was found to 
have a lower value.

Monte Carlo simulation of this model is difficult due to the sign problem
caused by the complex fermion determinant.
Among various reweighting-type methods,
the factorization method \cite{Anagnostopoulos:2001yb}
turned out to be particularly useful in the present case.
Assuming that the SO(4) symmetry is spontaneously broken down
either to SO(2) or to SO(3), the results of the GEM
(\ref{eq:previous_result}) and (\ref{eq:previous_result_SO3})
were reproduced \cite{Anagnostopoulos:2010ux,Anagnostopoulos:2011cn}.
However, the calculation of the free energy difference
had
large uncertainties, 
and it was not possible to determine which
vacuum is actually realized using this approach.

\section{Application of the CLM to 
the SO(4)-symmetric matrix model
\label{sec:Application-of-the}}

In this section, we explain how we apply the CLM 
to the SO(4)-symmetric matrix model (\ref{eq:partition_func}).
Including the symmetry breaking term (\ref{eq:boson_action_bdeform}),
we can write the partition function as
\begin{alignat}{3}
Z & =\int dX\,  w (X) \ , \quad \quad
w(X) =\left( \det D\right)^{N_{\rm f}}
e^{-(S_{{\rm b}}+\Delta S_{{\rm b}}) } \ .
\label{part-fn-with-det-rewrite}
\end{alignat}
The drift term that appears in the Langevin equation 
is given by
\begin{alignat}{3}
(v_\mu)_{ij} & =
\frac{1}{w(X)} \frac{\partial w(X)}{\partial (X_\mu)_{ji}} \\
&= 
- N \, \left(1+\varepsilon m_\mu \right) (X_\mu)_{ij} 
+ N_{\rm f} \,  (D^{-1})_{i\alpha , j\beta} (\Gamma_{\mu})_{\beta \alpha}
\label{drift-term}
\end{alignat}
as a function of the Hermitian matrices $X_\mu$.
Note that the second term in (\ref{drift-term}) 
is not Hermitian in general 
corresponding to the fact that the fermion determinant
is complex.
Thus, the application of the idea of stochastic quantization
naturally leads us to
complexifying the dynamical variables, which amounts
to regarding the Hermitian matrices $X_\mu$ as general complex matrices $X_\mu$.
Accordingly, the definition of the drift term (\ref{drift-term}) 
is extended to general complex matrices $X_\mu$ by analytic continuation.
Then we consider the fictitious-time evolution of
the general complex matrices $X_\mu$
described by the discretized version of the complex Langevin equation
\begin{equation}
X^{(\eta)}_\mu(t+\Delta t) = X^{(\eta)}_\mu(t) +
\Delta t \, v_\mu (X^{(\eta)}(t))
+ \sqrt{\Delta t} \, \eta_\mu (t) \ ,
\label{CLE-discretized}
\end{equation}
where $\eta_\mu (t)$ is 
an $N\times N$ Hermitian matrix generated with the probability
proportional to
$\ee^{-\frac{1}{4} \sum_t \, \tr \{ \eta_\mu (t)^2  \} } $.
The expectation values of the observables (\ref{lambda-def}) 
can be calculated as
\begin{alignat}{3}
\langle \lambda_\mu \rangle
&= \lim_{T\rightarrow \infty} 
\frac{1}{T}
\int_{t_0}^{t_0+T} \frac{1}{N}{\rm tr} \,  
\Big( X^{(\eta)}_{\mu}(t) \Big)^2  \ ,
\label{lambda-vev}
\end{alignat}
where $t_0$ represents the time required for thermalization and
$T$ should be large enough to achieve good statistics.

In order to justify the CLM, the probability distribution of the drift term
(\ref{drift-term}) measured during the complex Langevin simulation
should fall off exponentially or faster
at large magnitude \cite{Nagata:2016vkn}.
In the present model, this condition can be violated for two reasons.
First, the first term in (\ref{drift-term}) can be large
when the configuration $X^{(\eta)}_\mu(t)$ becomes too far from 
Hermitian.
Second, the second term in (\ref{drift-term}) can be large
when the Dirac operator $D$ has an eigenvalue close to zero.

In order to avoid the first problem, we use the gauge cooling \cite{Seiler:2012wz}.
Note that the original theory (\ref{part-fn-with-det-rewrite})
has the symmetry $X_\mu \mapsto g \, X_\mu \, g^{-1}$ with
$g \in {\rm SU}(N)$, under which
the drift term (\ref{drift-term}) transforms covariantly 
as $v_\mu \mapsto g \, v_\mu \, g^{-1}$
and the observables (\ref{lambda-def}) are invariant.
Upon complexifying the variables, the symmetry property of the
drift term and the observables enhances to
$X_\mu \mapsto g \, X_\mu \, g^{-1}$
with $g \in {\rm SL}(N,\bbC)$.
Using this fact, we can implement 
the gauge cooling procedure \cite{Seiler:2012wz}
in the Langevin process as
\begin{alignat}{3}
\tilde{X}^{(\eta)}_\mu(t) &= g \, X^{(\eta)}_\mu(t) \, g^{-1} \ ,
\label{CLE-cool-1}
\\
X^{(\eta)}_\mu(t+\Delta t) &= \tilde{X}^{(\eta)}_\mu(t) +
\Delta t
\, v_\mu (\tilde{X}^{(\eta)}(t))
+ 
\sqrt{\Delta t}
\, \eta(t) \ ,
\label{CLE-cool-2}
\end{alignat}
where the transformation matrix $g \in {\rm SL}(N,\bbC)$ is chosen
appropriately as a function of the configuration $X^{(\eta)}_\mu(t)$
before gauge cooling.
(See refs.~\cite{Nagata:2015uga,Nagata:2016vkn} 
for explicit justification.)

In order to keep the matrices $X^{(\eta)}_\mu(t)$ close to Hermitian,
%
we define the Hermiticity norm 
\begin{equation}
{\cal N}_{\rm H}=\frac{1}{4N}\sum_{\mu=1}^{4}{\rm tr}
\left[\left(X_{\mu}-X_{\mu}^{\dagger}\right)
\left(X_{\mu}-X_{\mu}^{\dagger}\right)^{\dagger}\right] \ ,
\label{eq:Hermitian_norm}
\end{equation}
which measures the deviation of $X_{\mu}$ from a Hermitian configuration,
and choose the ${\rm SL}(N,\bbC)$ transformation 
$g$ in (\ref{CLE-cool-1}) in such a way that
the norm is minimized. In practice, this is done 
by using the steepest descent method as follows.

\begin{figure}[t]
\centering{}
\includegraphics[width=8cm]{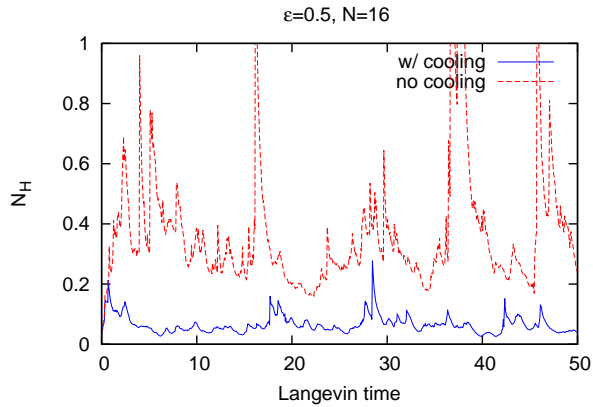}
\caption{(Left) The history of the 
Hermiticity norm \eqref{eq:Hermitian_norm} 
measured during the Langevin simulation
for $\varepsilon=0.5$
and $N=16$.
The solid line represents the case
with gauge cooling and the dashed line represents 
the case without gauge cooling. 
\label{fig:Hermitian-norm}}
\end{figure}

Let us consider an infinitesimal ${\rm SL}(N,\bbC)$ transformation
\begin{alignat}{3}
g = 1 + \epsilon_a t_a \ ,
\label{g-infinitesimal}
\end{alignat}
where $N\times N$ traceless
Hermitian matrices $t_a$ are the generators of ${\rm SU}(N)$ normalized as
$\tr (t_a t_b) =\delta_{ab}$.
Since the norm (\ref{eq:Hermitian_norm})
is invariant under ${\rm SU}(N)$, we restrict
the infinitesimal parameters $\epsilon_a$ to be real.
Under the infinitesimal transformation, 
we have
\begin{alignat}{3}
X_\mu & \mapsto X_\mu + \epsilon_a [t_a , X_\mu ] \ ,\nn \\
X_\mu^\dag & \mapsto X_\mu^\dag - \epsilon_a [t_a , X_\mu^\dag ] \ .
\label{X-tr-infinitesimal}
\end{alignat}
Therefore, the change of the Hermiticity norm (\ref{eq:Hermitian_norm})
becomes
\begin{alignat}{3}
\Delta {\cal N}_{\rm H} &=
\frac{1}{N} \epsilon_a  \sum_{\mu} \tr \Big(t_a [X_\mu , X_\mu^\dag ] \Big) \ ,
\label{norm-tr-infinitesimal}
\end{alignat}
from which the gradient of the norm is obtained as
\begin{alignat}{3}
f_a &= \frac{1}{N}
 \sum_{\mu} \tr \Big(t_a [X_\mu , X_\mu^\dag ] \Big) \ .
\label{gradient-norm}
\end{alignat}
Using this $f_a$, we consider a finite ${\rm SL}(N,\bbC)$ transformation
\begin{alignat}{3}
g = \ee^{- \alpha \, f_a t_a} \ ,
\label{g-finite}
\end{alignat}
where the real positive parameter $\alpha$ is chosen in such a way that
the Hermiticity norm (\ref{eq:Hermitian_norm}) is approximately minimized.
We repeat this procedure until the norm (\ref{eq:Hermitian_norm})
stops decreasing within certain accuracy.
%

In Fig.~\ref{fig:Hermitian-norm},
we plot the history of the Hermiticity norm \eqref{eq:Hermitian_norm}
measured during the Langevin simulation
for $\varepsilon =0.5$ and $N=16$.
Here and henceforth, 
the parameters $m_\mu$ in the SO(4)-breaking term
(\ref{eq:boson_action_bdeform})
are chosen as
\begin{equation}
(m_1,m_2,m_3,m_4)=(1,2,4,8) \ ,
\label{eq:boson_mass_used}
\end{equation}
and the Langevin step-size is chosen as $\Delta t=2.0\times10^{-4}$
unless stated otherwise.
We find that the gauge cooling keeps the Hermiticity norm
well under control.

\begin{figure}[t]
\centering{}
\includegraphics[width=7cm]{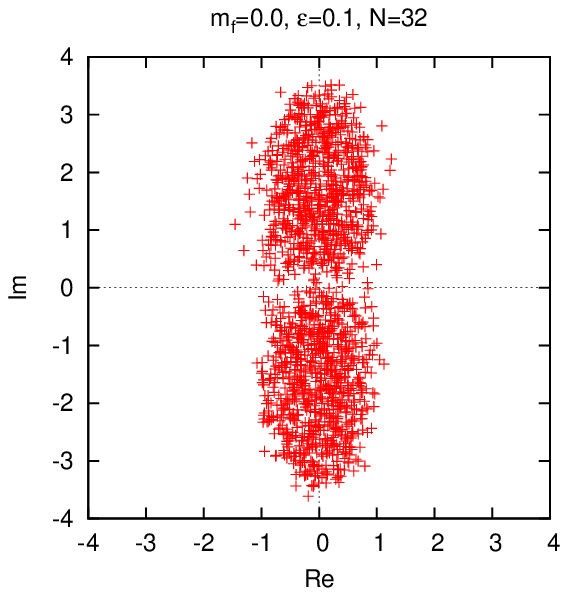}
\includegraphics[width=7cm]{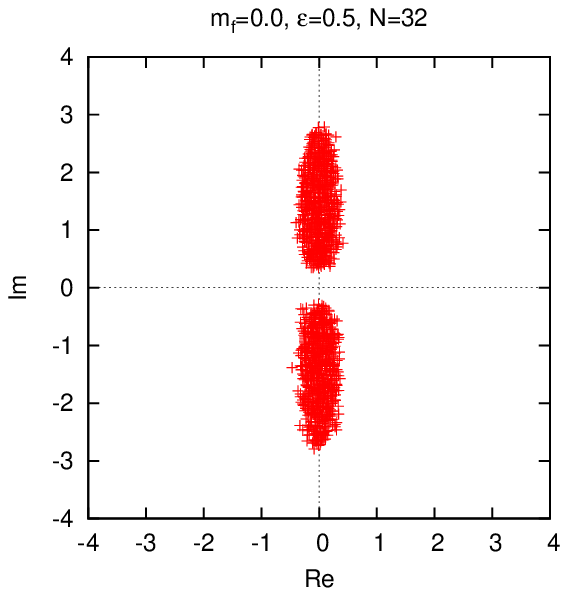}
\caption{The scatter plot for the eigenvalues of the Dirac operator
obtained during the complex Langevin simulation
of the undeformed model (\ref{part-fn-with-det-rewrite})
for $\varepsilon=0.1$ (Left) and $\varepsilon=0.5$ (Right)
with $N=32$.
\label{fig:ev-distribut}}
\end{figure}

Next we turn to the second problem, which 
is associated with the eigenvalues of
the Dirac operator $D$ close to zero.
In Fig.~\ref{fig:ev-distribut}, we plot the eigenvalue distribution of the
Dirac operator obtained during the complex Langevin simulation
for $\varepsilon=0.1$ (Left) and $\varepsilon=0.5$ (Right)
with $N=32$.
We find that there are many eigenvalues close to zero for $\varepsilon=0.1$,
but not for $\varepsilon=0.5$. 
This suggests that there is some critical $\varepsilon$, below which
the results of the CLM cannot be trusted because of the singular-drift
problem.
It turns out that the extrapolation to $\varepsilon =0$ is rather difficult
in this situation.

\begin{figure}[t]
\centering{}
\includegraphics[width=7cm]{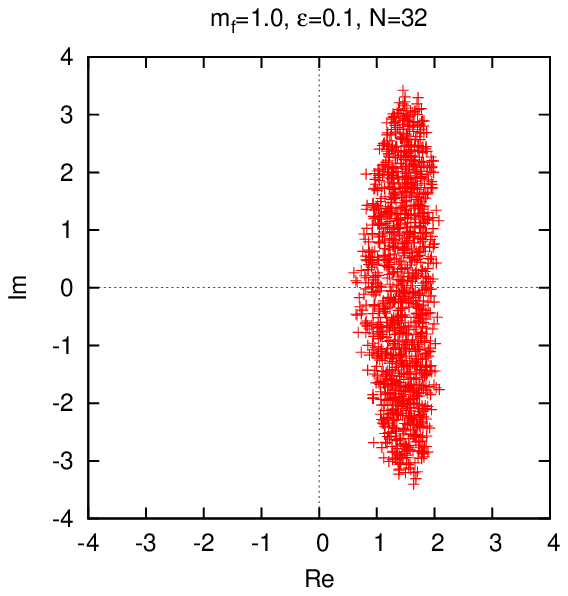}
\includegraphics[width=7cm]{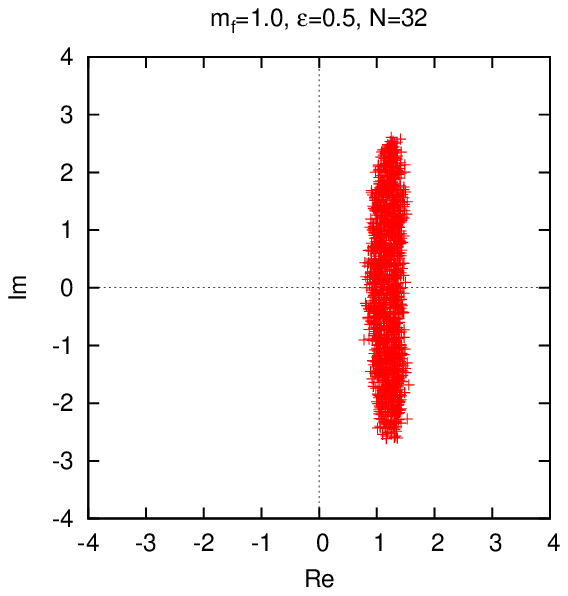}
\caption{The scatter plot for the eigenvalues of the Dirac operator
obtained during the complex Langevin simulation
of the deformed model 
defined by (\ref{part-fn-with-det-rewrite-deformed}) and (\ref{eq:4-th_mass})
for $\varepsilon=0.1$ (Left) and $\varepsilon=0.5$ (Right)
with $m_{{\rm f}}=1.0$ and $N=32$.
\label{fig:ev-distrib_shift}}
\end{figure}

In order to avoid this problem,
we add a fermion bilinear term
\begin{alignat}{3}
\Delta S_{{\rm f}} & =  -N 
\sum_{f=1}^{N_{\rm f}}
\sum_{\mu=1}^{4} M_\mu
\sum_{\alpha,\beta=1}^{2} 
\bar{\psi}_{\alpha}^{(f)}
\left(\Gamma_{\mu}\right)_{\alpha\beta} \, 
\psi_{\beta}^{(f)} 
\label{eq:fermion_action-deform}
\end{alignat}
to the action (\ref{eq:fermion_action}).
The partition function of the deformed model is defined as
\begin{alignat}{3}
& \tilde{Z}  =\int dX\,  \tilde{w} (X) \ , \quad \quad
\tilde{w}(X) =\left( \det \tilde{D} \right)^{N_{\rm f}}
e^{-(S_{{\rm b}}+\Delta S_{{\rm b}}) } \ ,
\nn \\
& \tilde{D}_{i\alpha , j \beta} 
 = \sum_{\mu=1}^{4} (\Gamma_{\mu})_{\alpha\beta} 
\Big(  (X_{\mu})_{ij} + M_\mu \delta_{ij} \Big) \ .
\label{part-fn-with-det-rewrite-deformed}
\end{alignat}
Note that the extra fermion bilinear term explicitly breaks the SO(4) symmetry
of the original model (\ref{eq:partition_func}).
Here we choose the parameters
$M_\mu$ in such a way that the SO(4) symmetry is broken
minimally.
Taking account of the ordering (\ref{lambda-ordering}),
we can preserve an SO(3) symmetry at $\varepsilon = 0$
by choosing
\begin{equation}
M_\mu=\left(0,0,0,m_{{\rm f}}\right) \ .
\label{eq:4-th_mass}
\end{equation}
We can then ask whether the SO(3) symmetry of this deformed model
is spontaneously broken in the large-$N$ limit.

In Fig.~\ref{fig:ev-distrib_shift},
we plot the eigenvalue
distribution of 
the Dirac operator (\ref{part-fn-with-det-rewrite-deformed})
obtained during the complex Langevin simulation of the deformed model
for $\varepsilon=0.1$ (Left) and $\varepsilon=0.5$ (Right)
with $m_{{\rm f}}=1.0$ and $N=32$.
We find that the distribution is shifted
in the real direction.
This is understandable since, at large $m_{{\rm f}}$,
the eigenvalue distribution of the Dirac operator would be 
distributed around $m_{{\rm f}}$.
As a result,
the distribution avoids the singularity even for $\varepsilon=0.1$
in contrast to the undeformed ($m_{{\rm f}}=0$) case. 
Therefore, we can extrapolate
$\varepsilon$ to zero using data obtained with
smaller $\varepsilon$ for finite $m_{{\rm f}}$. 
Eventually, we extrapolate the deformation parameter $m_{{\rm f}}$ to zero,
and compare the results with the prediction 
(\ref{eq:previous_result})
obtained by the GEM for the original model.

\begin{figure}[t]
\centering{}
\includegraphics[width=8cm]{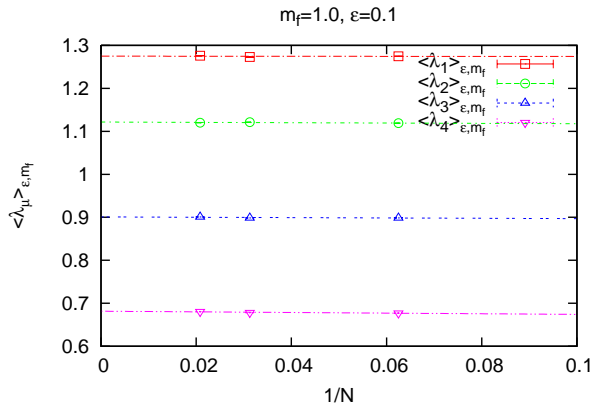}
\caption{The expectation values 
$\langle \lambda_{\mu} \rangle_{\varepsilon , m_{\rm f}}$ ($\mu=1,2,3,4$)
for the deformed model
defined by (\ref{part-fn-with-det-rewrite-deformed}) and (\ref{eq:4-th_mass})
are plotted against $1/N$
for $\varepsilon=0.1$ and $m_{{\rm f}}=1.0$. 
The straight lines represent fits
to the behavior $a + b/N$.
\label{fig:large-N}}
\end{figure}

\section{Results of our analysis
\label{sec:results}}

In this section, we present our results obtained by the CLM
as described in the previous section.
Let us recall that we have introduced an O($\varepsilon$) mass term 
(\ref{eq:boson_action_bdeform}) for the bosonic matrices,
which breaks the SO(4) symmetry explicitly.
In order to probe the SSB, we need to take the large-$N$
limit with fixed $\varepsilon$, and then make an extrapolation to 
$\varepsilon =0$.

In Fig.~\ref{fig:large-N},
the expectation values 
$\langle \lambda_{\mu} \rangle_{\varepsilon , m_{\rm f}}$ ($\mu=1,2,3,4$)
obtained for $N=16$, 32, 48 with $\varepsilon=0.1$ and $m_{{\rm f}}=1.0$
are plotted against $1/N$, where the data can be fitted
nicely to straight lines.
Thus we can 
extrapolate the expectation values to $N=\infty$
for each $\varepsilon$ and $m_{{\rm f}}$.
In what follows, we assume that the large-$N$ limit is already
taken in this way.

Next we would like to make an extrapolation to $\varepsilon =0$.
For that purpose, it is convenient to consider the ratio
\begin{alignat}{3}
\rho_{\mu}(\varepsilon ,m_{\rm f}) =
\frac{ \langle \lambda_{\mu} \rangle_{\varepsilon ,m_{\rm f}}}
{ \sum_{\nu=1}^4 \langle \lambda_{\nu} \rangle_{\varepsilon ,m_{\rm f}}}  \ .
\label{rho-def}
\end{alignat}
This is motivated from the fact that the mass term
(\ref{eq:boson_action_bdeform}) tends to make 
all the expectation values
$\langle \lambda_{\mu} \rangle_{\varepsilon ,m_{\rm f}}$
smaller than the value to be obtained in the $\varepsilon \rightarrow 0$ limit.
By taking the ratio (\ref{rho-def}), the finite $\varepsilon$ effects
are canceled by the denominator, 
and the extrapolation to $\varepsilon =0$ becomes easier.
Since $\varepsilon$ is a parameter in the action
(\ref{eq:boson_action_bdeform}),
the expectation values 
$\langle \lambda_{\mu} \rangle_{\varepsilon ,m_{\rm f}}$
and hence the ratios (\ref{rho-def})
can be expanded in a power series with respect to $\varepsilon$.
By taking the ratios,
the coefficients of higher order terms 
become smaller, and 
the truncation of the series becomes valid for 
a wider range of $\varepsilon$.

\begin{figure}[t]
\centering{}
\includegraphics[width=7cm]{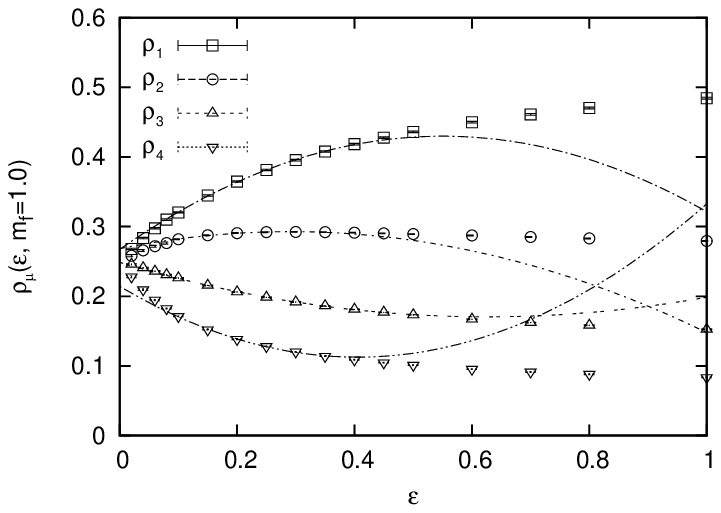}
\includegraphics[width=7cm]{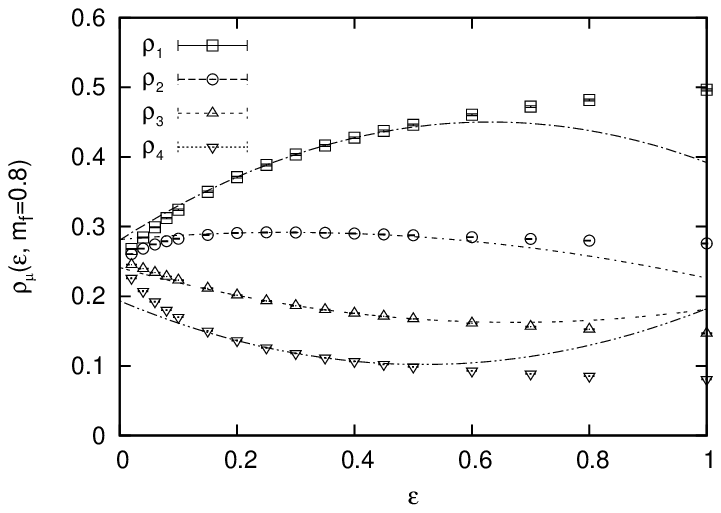}

\includegraphics[width=7cm]{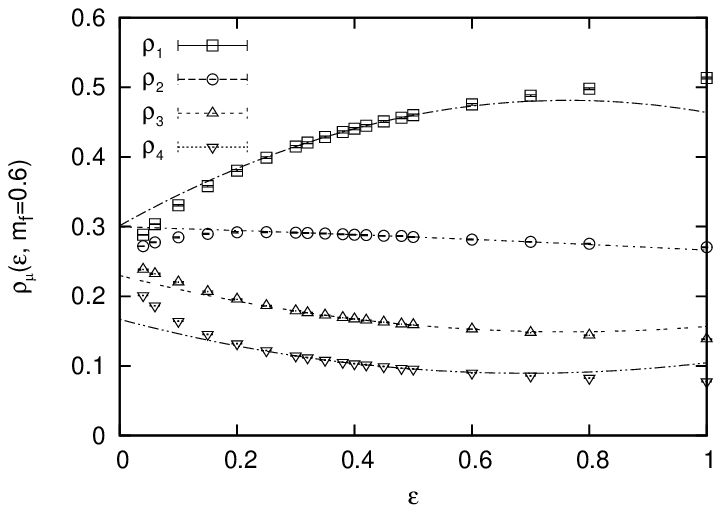}
\includegraphics[width=7cm]{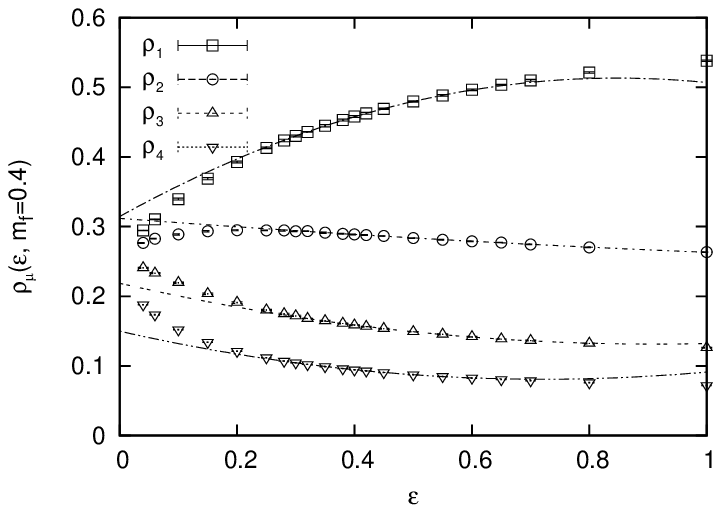}

\caption{The ratios $\rho_{\mu} (\varepsilon ,m_{\rm f})$
obtained after taking the large-$N$ limit
for the deformed model
defined by (\ref{part-fn-with-det-rewrite-deformed}) and (\ref{eq:4-th_mass})
are plotted against $\varepsilon$
for $m_{{\rm f}}=1.0$ (Top-Left),
0.8 (Top-Right), 0.6 (Bottom-Left) and 0.4 (Bottom-Right).
The lines represent fits 
to the quadratic form $a + b\varepsilon+c\varepsilon^{2}$.
\label{fig:extrolate_eps}}
\end{figure}

In Fig.~\ref{fig:extrolate_eps},
we plot the ratio (\ref{rho-def})
against $\varepsilon$ for $m_{{\rm f}}=1.0$ (Top-Left),
0.8 (Top-Right), 0.6 (Bottom-Left) and 0.4 (Bottom-Right).
The data obtained at small $\varepsilon$ suffer from
the singular-drift problem, and hence cannot be trusted.
Here the condition for justifying the CLM 
proposed recently in ref.~\cite{Nagata:2016vkn}
turns out to be useful since it enables us to 
determine the range of validity
as we explain in appendix \ref{sec:region-of-validity}.
Taking this into account, we fit the data 
in Fig.~\ref{fig:extrolate_eps}
to the quadratic form using 
the fitting range given in Table \ref{tab:The-list-m4},
where we also present the extrapolated values.
We find for each value of $m_{\rm f}$ that
$\rho_1 (\varepsilon ,m_{\rm f})$
and $\rho_2 (\varepsilon ,m_{\rm f})$
approach the same value
in the $\varepsilon \rightarrow 0$ limit,
while the others approach smaller values.
This implies that the SSB from SO(3) to SO(2) occurs in the deformed model.

\begin{figure}[t]
\centering{}
\includegraphics[width=8cm]{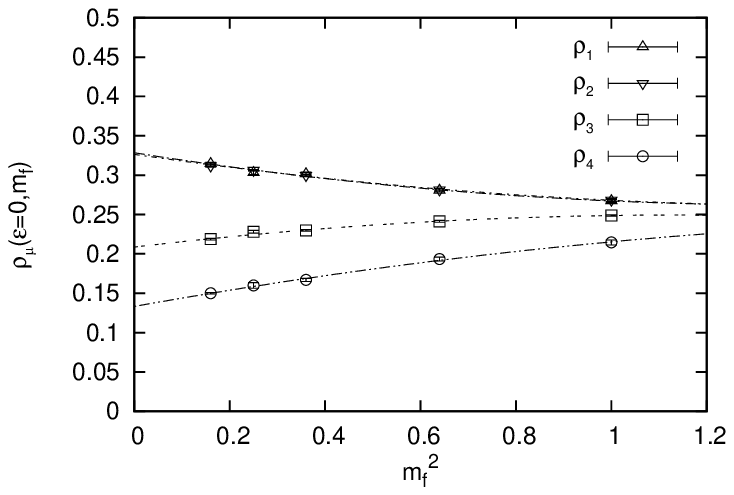}
\caption{The extrapolated values
$\lim_{\varepsilon \rightarrow 0}\rho_{\mu} (\varepsilon ,m_{\rm f})$ 
for the deformed model
defined by (\ref{part-fn-with-det-rewrite-deformed}) and (\ref{eq:4-th_mass})
are plotted against
$m_{\rm f}^2$. The lines represent fits 
to the quadratic form 
$a+b x + c x^2$ with $x=m_{\rm f}^2$
using the data within the region $0.4\leq m_{{\rm f}}\leq1.0$. 
\label{fig:extrapolate_mf}}
\end{figure}

In Fig.~\ref{fig:extrapolate_mf},
we plot the extrapolated values
$\lim_{\varepsilon \rightarrow 0} \rho_{\mu} (\varepsilon ,m_{\rm f})$
obtained in this way against $m_{\rm f}^2$.
We find that our results within $0.4 \le m_{{\rm f}} \le 1.0$ can be
nicely fitted to the quadratic behavior,
which is motivated by a 
power series expansion of the expectation values 
$\langle \lambda_{\mu} \rangle_{\varepsilon ,m_{\rm f}}$
with respect to $m_{\rm f}$.\footnote{The odd order terms in $m_{\rm f}$
do not appear due to the symmetry $m_{\rm f}\rightarrow -m_{\rm f}$
of the expectation values.}
Extrapolating $m_{{\rm f}}$ to zero, we obtain
$\lim_{m_{{\rm f}}\rightarrow0}
\lim_{\varepsilon\rightarrow0}
\rho_{\mu}(\varepsilon,m_{{\rm f}})
=0.328(4), 0.326(2), 0.208(2), 0.133(2)$
for $\mu=1,2,3,4$,
which shows 
%
that the SO(4) symmetry of 
the undeformed model ($m_{\rm f} = 0$)
is spontaneously broken down to SO(2). 
Moreover, using an exact result 
$\sum_{\mu=1}^4 \langle \lambda_\mu \rangle = 4 + 2r = 6 $ \cite{Nishimura:2001sq}
for the present $r=1$ case, we obtain
\begin{equation}
\left\langle \lambda_{1}\right\rangle =1.97(2)\ ,\quad
\left\langle \lambda_{2}\right\rangle =1.96(1)\ ,\quad
\left\langle \lambda_{3}\right\rangle =1.25(1)\ ,\quad
\left\langle \lambda_{4}\right\rangle =0.80(1)\ ,
\label{eq:deform-4th_result}
\end{equation}
which agree well with the results \eqref{eq:previous_result} 
obtained by the GEM.
Here we emphasize that in the GEM, 
the true vacuum was determined by comparing the free energy
obtained for the SO($2$) vacuum and the SO($3$) vacuum.
In contrast, the CLM enables us to determine the true vacuum directly 
without having to compare the free energy for different vacua.

As a further consistency check,
we repeat the same analysis
with a different choice of the deformation parameter
$M_{\mu}=\left(0,0,m_{{\rm f}},0\right)$
in (\ref{part-fn-with-det-rewrite-deformed}) 
instead of (\ref{eq:4-th_mass}).
We find that
the results obtained after the extrapolation $m_{{\rm f}}\rightarrow 0$
turn out to be consistent with the ones obtained above.
See appendix~\ref{sec:3rd-fermion-case}
for the details.


\begin{table}[H]
\centering{}
\begin{tabular}{|c|c|c|c|}
\hline 
$m_{{\rm f}}$  & $\mu$  & fitting range  & extrapolated value\tabularnewline
\hline 
\hline 
\multirow{4}{*}{1.0} & 1  & $0.1\leq\varepsilon\leq0.4$  & 0.2673(20)\tabularnewline
\cline{2-4} 
 & 2  & $0.1\leq\varepsilon\leq0.35$  & 0.2685(18)\tabularnewline
\cline{2-4} 
 & 3  & $0.1\leq\varepsilon\leq0.50$  & 0.2487(09)\tabularnewline
\cline{2-4} 
 & 4  & $0.1\leq\varepsilon\leq0.35$  & 0.2144(32)\tabularnewline
\hline 
\multirow{4}{*}{0.8} & 1  & $0.2\leq\varepsilon\leq0.4$  & 0.2806(21)\tabularnewline
\cline{2-4} 
 & 2  & $0.2\leq\varepsilon\leq0.4$  & 0.2815(13)\tabularnewline
\cline{2-4} 
 & 3  & $0.2\leq\varepsilon\leq0.5$  & 0.2413(11)\tabularnewline
\cline{2-4} 
 & 4  & $0.2\leq\varepsilon\leq0.4$  & 0.1934(24)\tabularnewline
\hline 
\multirow{4}{*}{0.6} & 1  & $0.3\leq\varepsilon\leq0.5$  & 0.3014(24)\tabularnewline
\cline{2-4} 
 & 2  & $0.3\leq\varepsilon\leq0.7$  & 0.2997(13)\tabularnewline
\cline{2-4} 
 & 3  & $0.3\leq\varepsilon\leq0.6$  & 0.2298(10)\tabularnewline
\cline{2-4} 
 & 4  & $0.3\leq\varepsilon\leq0.5$  & 0.1669(19)\tabularnewline
\hline 
\multirow{4}{*}{0.4} & 1  & $0.3\leq\varepsilon\leq0.6$  & 0.3144(20)\tabularnewline
\cline{2-4} 
 & 2  & $0.3\leq\varepsilon\leq0.7$  & 0.3125(11)\tabularnewline
\cline{2-4} 
 & 3  & $0.3\leq\varepsilon\leq0.8$  & 0.2183(07)\tabularnewline
\cline{2-4} 
 & 4  & $0.3\leq\varepsilon\leq0.6$  & 0.1495(08)\tabularnewline
\hline 
\end{tabular}
\caption{The fitting range 
used in Fig.~\ref{fig:extrolate_eps}
for the $\varepsilon \rightarrow 0$ extrapolations
is listed with the extrapolated values obtained by the fits.
\label{tab:The-list-m4}}
\end{table}

\section{Summary and discussion
\label{sec:Summary-and-discussion}
}

In this paper, we have shown that the CLM can be successfully applied
to a matrix model, in which the SSB of SO(4) is expected to occur due to
the phase of the complex fermion determinant.
The SSB does not occur if the phase is quenched,
which implies that it is extremely hard 
to investigate this phenomenon
by reweighting-based Monte Carlo methods. 
In the factorization method, for instance,
one introduces a constraint with 
some parameters and extremizes the free energy with respect to these
parameters. 
While this has been done successfully 
in refs.~\cite{Anagnostopoulos:2010ux,Anagnostopoulos:2011cn},
the comparison of the free energy for
the SO(2) and SO(3) vacua turns out to be subtle
and a definite conclusion on the true vacuum was not reached.
In contrast, we have shown by the CLM that the SSB
from SO(4) down to SO(2) occurs 
as predicted by the GEM.
%

For the success of the CLM,
it was crucial
to overcome the singular-drift problem associated with the appearance
of nearly zero eigenvalues of the Dirac operator.
The gauge cooling was used to suppress the excursions in the imaginary directions,
but the singular-drift problem in the present case
was too severe to be solved by the gauge cooling.
This is understandable because the fermionic variables are
exactly ``massless'' in the present case.
Our strategy to overcome the singular-drift problem
was to deform the Dirac operator in such a way that
the singular-drift problem is avoided while maintaining the qualitative
feature of the vacuum as much as possible.
On top of this, we have to introduce 
an O($\varepsilon$) 
symmetry breaking term to probe the SSB, which should be removed after
taking the large-$N$ limit.
In making the $\varepsilon\rightarrow 0$ extrapolations,
the criterion for correct convergence 
proposed in ref.~\cite{Nagata:2016vkn}
turns out to be useful since it tells us
the range of parameters for which
the CLM is free from the singular-drift problem 
and the results are trustable.
The order parameters 
obtained after extrapolating the deformation parameter
to zero turn out to be consistent
with the prediction by the GEM.

We have actually tried two types of deformation 
to avoid the singular-drift problem
and confirmed that the extrapolated results agree with each other
within fitting errors.
While this confirms the validity of the extrapolations to some extent,
we cannot exclude the possibility that something
dramatic happens when the deformation parameter approaches zero.
Let us recall, however, that the singular-drift problem can occur
at some point in the parameter space even if the system
itself does not undergo any dramatic change.
For instance, in QCD at finite density, 
the singular-drift problem is anticipated 
to occur at the quark chemical potential 
$\mu \gtrsim m_{\pi}/2$, where $m_{\pi}$ is the pion mass, but
the first order transition to the phase of nuclear matter
occurs at $\mu \sim m_{\rm N}/3$, where
$m_{\rm N}$ is the nucleon mass.
Nothing really happens
in the wide parameter range 
$0 \lesssim \mu \lesssim m_{\rm N}/3$.
This example clearly shows that 
the singular-drift problem 
has more to do with the methodology rather than the physics of the
system to be investigated.
%
%
%

The CLM with the proposed strategy can be directly applied
to the type IIB matrix model, which is conjectured to be 
a nonperturbative formulation of type IIB superstring theory 
in ten dimensions \cite{Ishibashi:1996xs}.
While the SO(10) symmetry
of the model is expected to be 
spontaneously broken down to SO(4) for consistency
with our 4d space-time,
the GEM predicts that 
it
is spontaneously
broken down to SO(3) rather than SO(4) \cite{Nishimura:2011xy}.
It would be interesting to investigate this issue using the CLM extending
the present work.

We consider that the same strategy would be useful also in applying the 
CLM to finite density QCD at low temperature and high density,
where various exotic condensates 
are speculated to form \cite{Rajagopal:2000wf}
due to the complex fermion determinant. 
In this case, one can deform the Dirac operator by
switching on the corresponding fermion bilinear term without 
disturbing
the vacuum significantly. 
Now that we have a useful criterion \cite{Nagata:2016vkn}
for justifying the CLM,
we can try possible deformations
and see whether any of them allows us to extrapolate
the deformation parameter to zero
within the region of validity.

\section*{Acknowledgements}

The authors would like to 
thank 
K.N.~Anagnostopoulos, T.~Azuma, 
K.~Nagata, S.K.~Papadoudis and S.~Shimasaki 
for valuable discussions.
Y.~I.\ is supported by JICFuS.
J.~N.\ is supported in part by Grant-in-Aid 
for Scientific Research (No.\ 23244057 and 16H03988)
from Japan Society for the Promotion of Science.


\appendix

\begin{figure}[H]
\centering{}
\includegraphics[width=7cm]{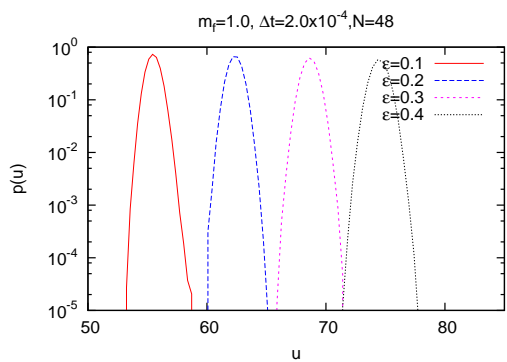}
\caption{The probability distribution $p\left(u\right)$ of the
magnitude of the drift term $u$ is plotted 
in the log scale
for various $\varepsilon$ with $m_{{\rm f}}=1.0$ and $N=48$.
\label{fig:drift_hisotgram_1}}
\end{figure}

\begin{figure}[H]
\centering{}
\includegraphics[width=7cm]{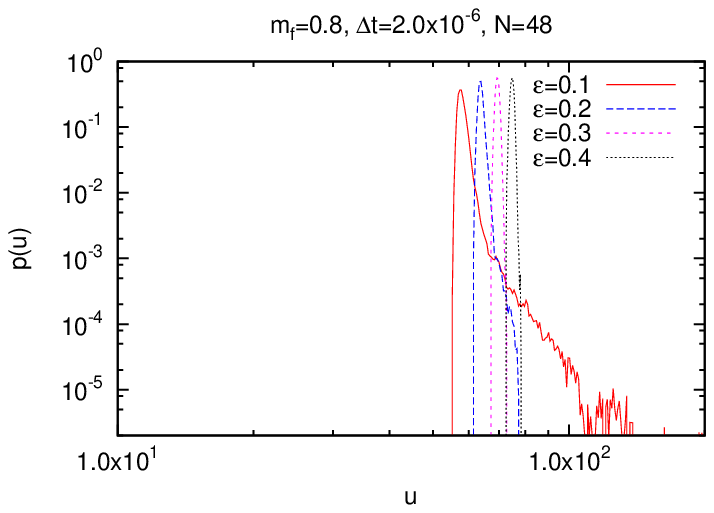}
\includegraphics[width=7cm]{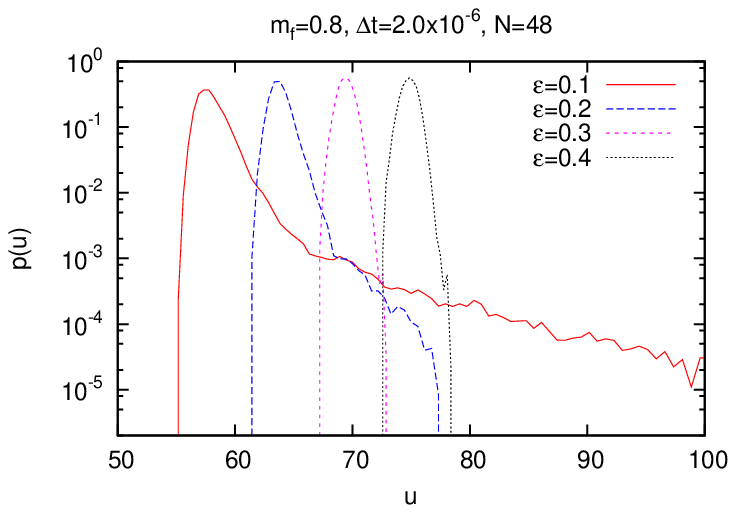}
\caption{The probability distribution $p\left(u\right)$ of the
magnitude of drift term $u$ is plotted for various 
$\varepsilon$ with $m_{{\rm f}}=0.8$ and $N=48$.
The step-size had to be lowered to $\Delta t=2.0\times10^{-6}$
in order to probe the behavior of the tail correctly.
A log-log plot (Left) and a semi log plot (Right) are shown.
\label{fig:drift_hisotgram_1-1}}
\end{figure}

\section{How to determine the region of validity}
\label{sec:region-of-validity}

In this appendix, we explain how to determine the region of
validity of the CLM. 
When the symmetry breaking parameter $\varepsilon$ becomes small,
the singular-drift problem occurs and the results obtained by the CLM
can no longer be trusted.
In order to make $\varepsilon \rightarrow 0$ extrapolations,
it is important to determine the value of $\varepsilon$, below which
the results become unreliable.
Here we use the criterion based on the 
argument for justifying the CLM \cite{Nagata:2016vkn}.
For that, we calculate the magnitude of the drift term for each configuration
and obtain its probability distribution. If the tail of the distribution
falls off exponentially or faster, 
we can trust the results obtained with those
simulation parameters.
We find that the finite step-size effects can 
modify the tail of the distribution significantly
without changing the expectation
values $\langle \lambda_{\mu} \rangle_{\varepsilon , m_{\rm f}}$.
In order to make the plots in this section, we 
therefore have to decrease the step-size 
when it turns out to be necessary.

Let us define the magnitude of the drift term by 
\begin{equation}
u= \sqrt{\frac{1}{4N}
\sum_{\mu=1}^{4}{\rm Tr}\left(v_{\mu}^{\dagger}v_{\mu}\right)} \ ,
\end{equation}
where $v_{\mu}$ is the drift term defined by (\ref{drift-term}).
Then, we define the probability distribution $p\left(u\right)$
with the normalization $\int_{0}^{\infty}du\, p\left(u\right)=1$.
In Fig.~\ref{fig:drift_hisotgram_1},
we plot $p\left(u\right)$ against $u$ in the log scale
for various $\varepsilon$
with $m_{{\rm f}}=1.0$ and $N=48$.
We find that $p\left(u\right)$ falls off exponentially or faster
for all the $\varepsilon$.
Thus, we can trust the results obtained in this region.

\begin{figure}[t]
\centering{}
\includegraphics[width=7cm]{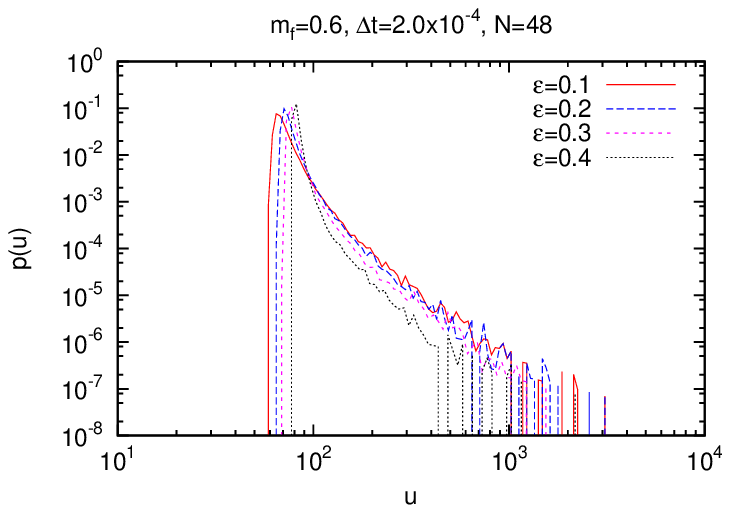}
\includegraphics[width=7cm]{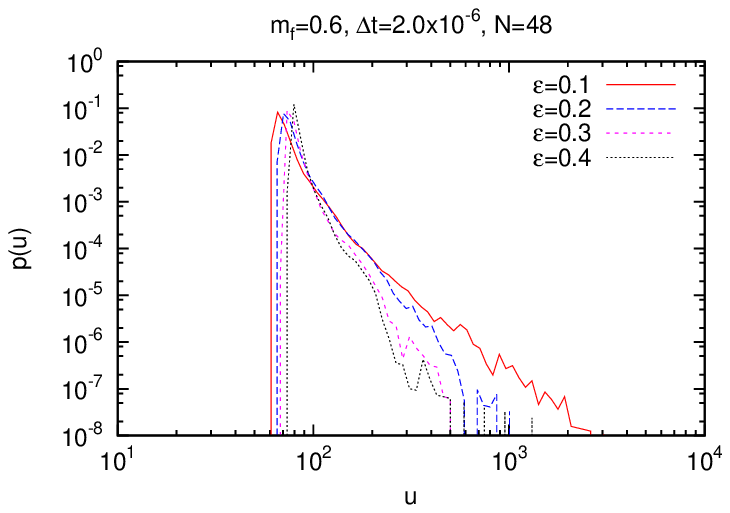}
\caption{The probability distribution $p\left(u\right)$ of the
magnitude of drift term $u$ is shown
for various $\varepsilon$ with
$m_{{\rm f}}=0.6$ and $N=48$ in log-log plots.
The Langevin step-size 
is chosen to be $\Delta t=2.0\times10^{-4}$ (Left)
and $2.0\times10^{-6}$ (Right).
\label{fig:drift_hisotgram_2}}
\end{figure}

\begin{figure}[t]
\centering{}
\includegraphics[width=7cm]{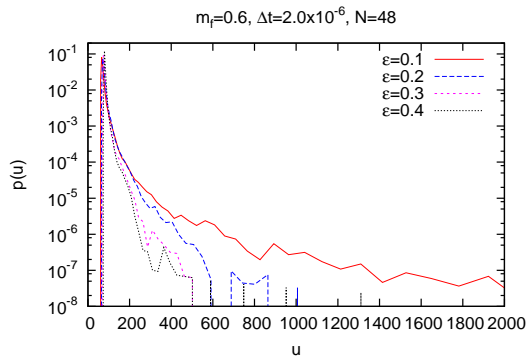}
\caption{A semi-log plot of the data 
in the right panel of Fig.~\ref{fig:drift_hisotgram_2}.
\label{fig:drift_hisotgram_2_1}}
\end{figure}

\begin{figure}[t]
\centering{}
\includegraphics[width=7cm]{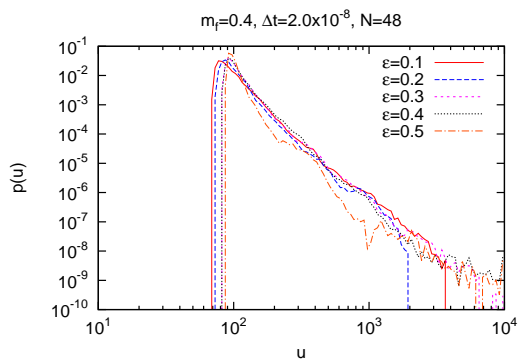}
\caption{The probability distribution $p\left(u\right)$ of the
magnitude of drift term $u$ is shown 
for various $\varepsilon$ with
$m_{{\rm f}}=0.4$ and $N=48$ in a log-log plot.
The Langevin step-size is chosen to be $\Delta t=2.0\times10^{-8}$.
\label{fig:drift_hisotgram_3}}
\end{figure}

In Fig.~\ref{fig:drift_hisotgram_1-1},
we show
a log-log plot (Left) and a semi log plot (Right)
of the distribution $p\left(u\right)$ for various $\varepsilon$
with $m_{{\rm f}}=0.8$ and $N=48$.
Since the drift term can become fairly large for $\varepsilon=0.1$,
we decrease the Langevin step-size to 
$\Delta t=2.0\times10^{-6}$ in order to probe the tail of the 
distribution correctly.
We find that the distribution falls off exponentially or faster
for $\varepsilon \ge 0.2$, but a power-law tail develops for 
$\varepsilon =0.1$.
Therefore, we can trust the data for $\varepsilon \ge 0.2$, but 
not the ones at $\varepsilon =0.1$.

In Fig.~\ref{fig:drift_hisotgram_2},
we show a log-log plot of 
$p\left(u\right)$ for various $\varepsilon$
with $m_{{\rm f}}=0.6$ and $N=48$.
Here the drift term tends to become even larger
than in the $m_{{\rm f}}=0.8$ case,
and we have to investigate the tail of the distribution
more carefully.
We therefore present the results obtained for
two Langevin step-size, $\Delta t=2.0\times10^{-4}$ (Left)
and $2.0\times10^{-6}$ (Right).
Indeed, we find that
the behavior of the tail
seems to change qualitatively by decreasing the step-size.
In Fig.~\ref{fig:drift_hisotgram_2_1}, we show a semi-log plot 
for $\Delta t=2.0\times10^{-6}$, which suggests that 
the tail of the distribution falls off exponentially
for $\varepsilon \ge 0.3$, but not for $\varepsilon = 0.1$.
The result for $\varepsilon = 0.2$ is marginal.
We may therefore trust the results for $\varepsilon \ge 0.3$.

In Fig.~\ref{fig:drift_hisotgram_3},
we show a log-log plot of 
$p\left(u\right)$ for various $\varepsilon$
with $m_{{\rm f}}=0.4$ and $N=48$.
Here we have decreased the Langevin step-size to
$\Delta t=2.0\times10^{-8}$, but
the tail of the distribution still
follows a power law for all values of $\varepsilon$ within the region.
However, 
the comparison of the two plots in Fig.~\ref{fig:drift_hisotgram_2}
suggests a possibility that the step-size $\Delta t$
should be decreased further to see the behavior of the tail correctly.
Thus for the $m_{{\rm f}}=0.4$ case alone,
we had to determine the lower end of the
fitting range empirically from the plausibility of the fit 
to the quadratic behavior.
Even if we omit the $m_{{\rm f}}=0.4$ point in
Fig.~\ref{fig:extrolate_eps},
the values obtained by extrapolations to $m_{{\rm f}}=0$
remain almost the same.

\section{Results for another type of the fermion bilinear term
\label{sec:3rd-fermion-case}}

In this appendix, we present the results obtained by
choosing the deformation parameters 
in (\ref{part-fn-with-det-rewrite-deformed}) as 
\begin{equation}
M_{\mu}=\left(0,0,m_{{\rm f}},0\right)
\label{eq:mass-deform_3rd_direction}
\end{equation}
instead of (\ref{eq:4-th_mass}).
Taking into account the ordering 
(\ref{lambda-ordering}),
we can preserve only an SO(2) symmetry with this choice.
%

\begin{figure}[t]
\begin{centering}
\includegraphics[width=7cm]{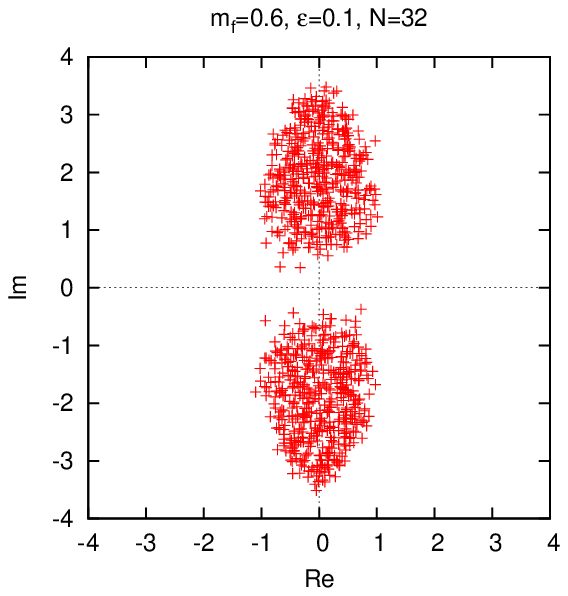}
\includegraphics[width=7cm]{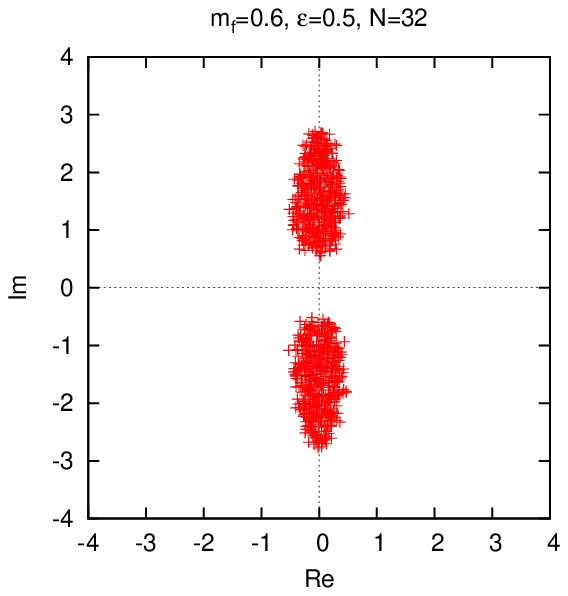}
\par\end{centering}

\caption{The scatter plot for the eigenvalues of the Dirac operator
obtained during the complex Langevin simulation
of the deformed model 
defined by (\ref{part-fn-with-det-rewrite-deformed}) and 
(\ref{eq:mass-deform_3rd_direction})
for $\varepsilon=0.1$ (Left)
and $\varepsilon=0.5$ (Right)
with $m_{{\rm f}}=0.6$ and $N=32$.
\label{fig:scattered-3rd-direction} }
\end{figure}

In Fig.~\ref{fig:scattered-3rd-direction},
we plot the eigenvalue
distribution of the Dirac operator \eqref{part-fn-with-det-rewrite-deformed}
for $\varepsilon=0.1$ (Left)
and $\varepsilon=0.5$ (Right)
with $m_{{\rm f}}=0.6$ and $N=32$.
We find that the distribution is separated 
in the imaginary direction.
This is understandable since, at large $m_{{\rm f}}$,
the eigenvalue distribution of the Dirac operator would be 
distributed around $\pm i \, m_{{\rm f}}$.
As a result, the singularity at the origin can be avoided
for even smaller $\varepsilon$ than in the case of \eqref{eq:4-th_mass}. 
This enables us to extrapolate
$\varepsilon$ to zero using the data obtained 
in the large-$N$ limit
for finite $m_{{\rm f}}$.

\begin{figure}[tp]
\centering{}
\includegraphics[width=7cm]{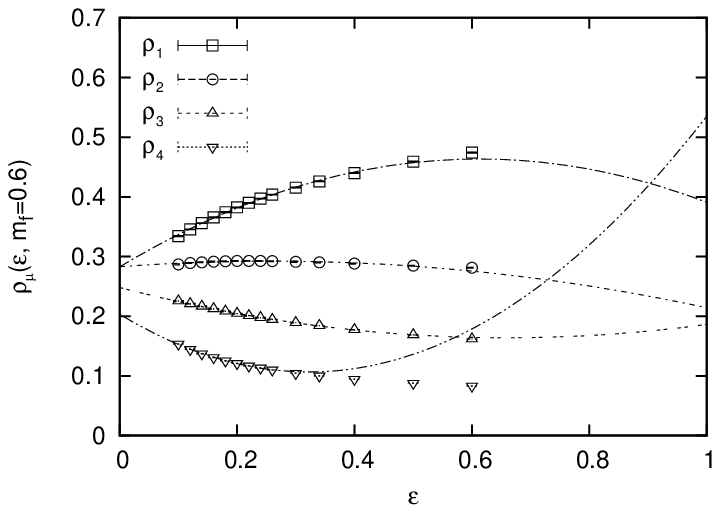}
\includegraphics[width=7cm]{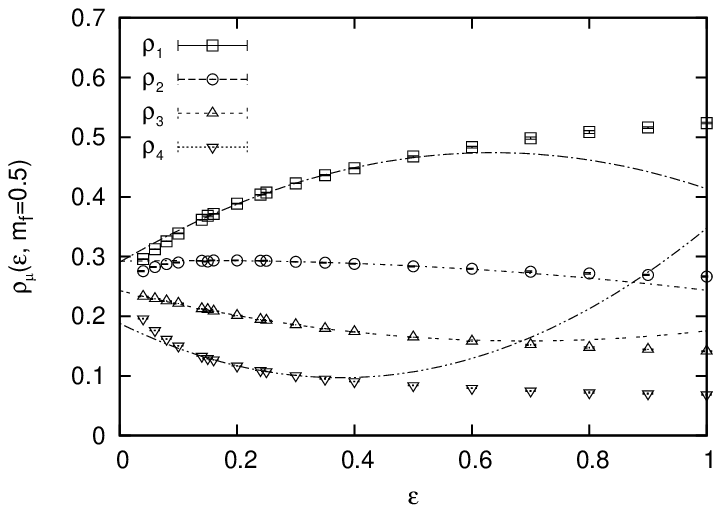}

\includegraphics[width=7cm]{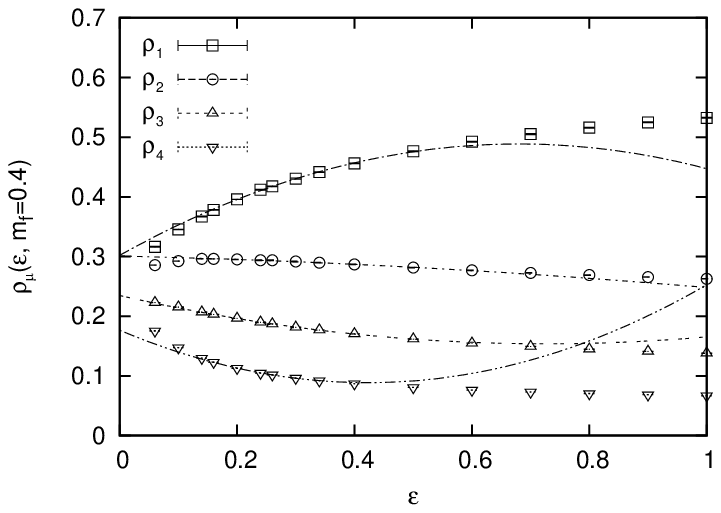}
\includegraphics[width=7cm]{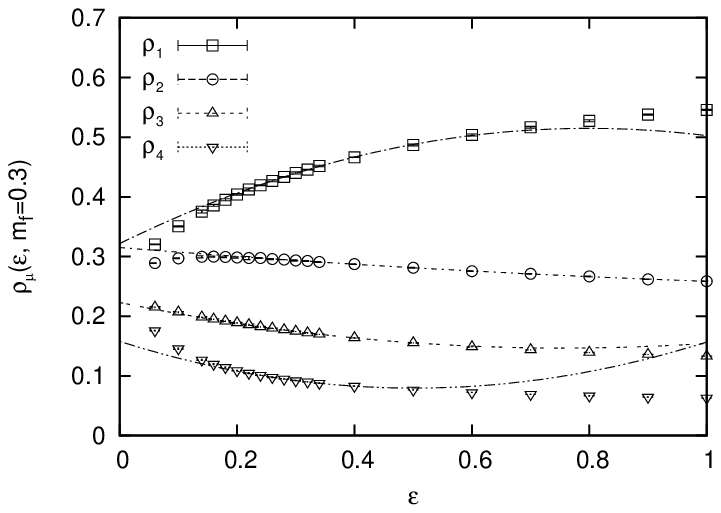}

\includegraphics[width=7cm]{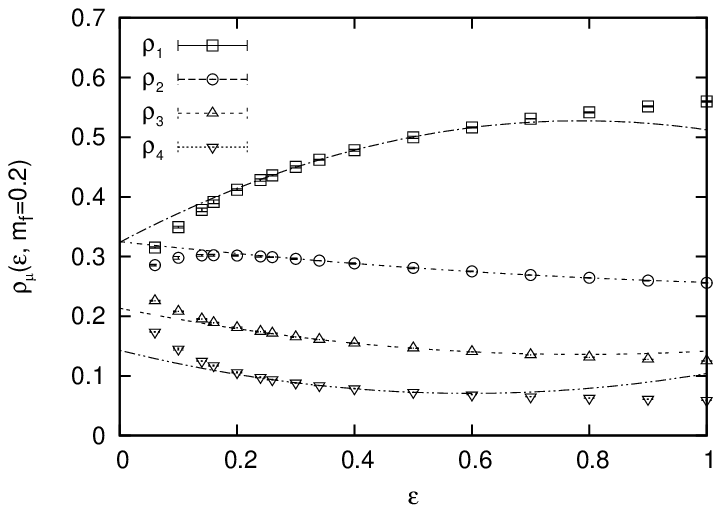}

\caption{
The ratios $\rho_{\mu} (\varepsilon ,m_{\rm f})$
obtained after taking the large-$N$ limit
for the deformed model
defined by (\ref{part-fn-with-det-rewrite-deformed}) and 
(\ref{eq:mass-deform_3rd_direction})
are plotted against $\varepsilon$
for $m_{{\rm f}}=0.6$ (Top-Left), 0.5 (Top-Right), 
0.4 (Middle-Left), 0.3 (Middle-Right) and 0.2 (Bottom).
The lines represent fits 
to the quadratic form $a + b\varepsilon+c\varepsilon^{2}$.
\label{fig:extrapolate_eps_3rd}}
\end{figure}

In Fig.~\ref{fig:extrapolate_eps_3rd}, we plot the ratios (\ref{rho-def})
obtained after taking the large-$N$ limit
against $\varepsilon$ 
for $m_{{\rm f}}=0.6$ (Top-Left), 0.5 (Top-Right), 
0.4 (Middle-Left), 0.3 (Middle-Right) and 0.2 (Bottom).
The data obtained for small $\varepsilon$
cannot be trusted because of the singular-drift problem.
We fit the data 
in Fig.~\ref{fig:extrapolate_eps_3rd}
to the quadratic form using 
the fitting range given in Table \ref{tab:The-list-m3},
where we also present the extrapolated values.
We find for each value of $m_{\rm f}$ that
$\rho_1 (\varepsilon ,m_{\rm f})$
and $\rho_2 (\varepsilon ,m_{\rm f})$
approach the same value
in the $\varepsilon \rightarrow 0$ limit,
while the others approach smaller values.

\begin{figure}[tp]
\centering{}
\includegraphics[width=8cm]{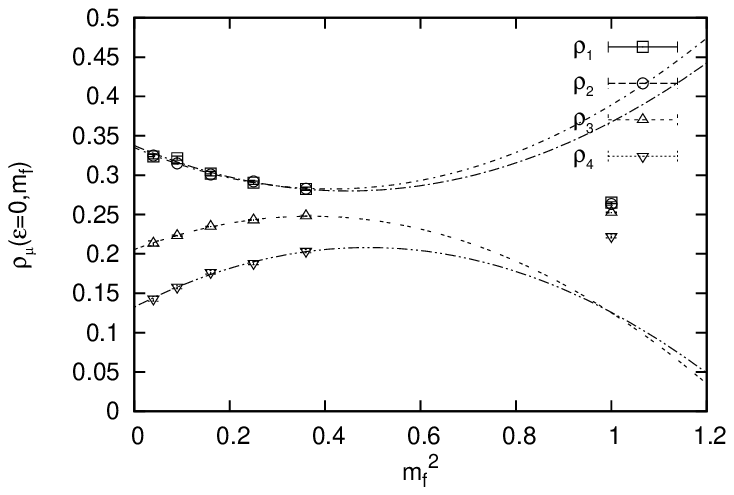}
\caption{The extrapolated values
$\lim_{\varepsilon \rightarrow 0}\rho_{\mu} (\varepsilon ,m_{\rm f})$
for the deformed model
defined by (\ref{part-fn-with-det-rewrite-deformed}) and 
(\ref{eq:mass-deform_3rd_direction})
are plotted against
$m_{\rm f}^2$. The lines represent fits 
to the quadratic form $a+b x+ x^2$ with $x=m_{\rm f}^2$
using the data within the region $ 0.2 \leq m_{{\rm f}}\leq 0.6$. 
\label{fig:extrapolate_mf_3rd}}
\end{figure}

In Fig.~\ref{fig:extrapolate_mf_3rd},
we plot the extrapolated values
$\lim_{\varepsilon \rightarrow 0} \rho_{\mu} (\varepsilon ,m_{\rm f})$
obtained in this way against $m_{{\rm f}}^2$. 
We find that our results within $0.2 \le m_{{\rm f}} \le 0.6$ can be
nicely fitted to the quadratic behavior.
Extrapolating $m_{{\rm f}}$ to zero, we obtain
$\lim_{m_{{\rm f}}\rightarrow 0}
\lim_{\varepsilon \rightarrow 0} \rho_{\mu} (\varepsilon ,m_{\rm f})
= 0.337(6)$, 0.335(2), 0.205(2), 0.132(4) for $\mu=1,2,3,4$.
Using an exact result 
$\sum_{\mu=1}^4 \langle \lambda_\mu \rangle 
= 4 + 2r = 6 $ \cite{Nishimura:2001sq}
for the present $r=1$ case, we obtain
\begin{alignat}{3}
\left\langle \lambda_1 \right\rangle =2.02(4) \ , \quad
\left\langle \lambda_2 \right\rangle =2.01(1) \ , \quad
\left\langle \lambda_3 \right\rangle =1.23(1) \ , \quad
\left\langle \lambda_4 \right\rangle =0.79(2) \ , \quad
\label{eq:deform-3rd_result}
\end{alignat}
%
%
which are consistent with the results \eqref{eq:deform-4th_result}
obtained with the choice (\ref{eq:4-th_mass}) for the deformation.
This supports the validity of our analysis.
%

\begin{table}[H]
\begin{centering}
\begin{tabular}{|c|c|c|c|}
\hline 
$m_{{\rm f}}$ & $\mu$ & fitting range & extrapolated value\tabularnewline
\hline 
\hline 
\multirow{4}{*}{0.6} & 1 & $0.1\leq\varepsilon\leq0.5$ & 0.2825(21)\tabularnewline
\cline{2-4} 
 & 2 & $0.1\leq\varepsilon\leq0.5$ & 0.2828(14)\tabularnewline
\cline{2-4} 
 & 3 & $0.1\leq\varepsilon\leq0.5$ & 0.2481(08)\tabularnewline
\cline{2-4} 
 & 4 & $0.1\leq\varepsilon\leq0.26$ & 0.2035(22)\tabularnewline
\hline 
\multirow{4}{*}{0.5} & 1 & $0.14\leq\varepsilon\leq0.5$ & 0.2904(21)\tabularnewline
\cline{2-4} 
 & 2 & $0.14\leq\varepsilon\leq0.6$ & 0.2921(12)\tabularnewline
\cline{2-4} 
 & 3 & $0.14\leq\varepsilon\leq0.5$ & 0.2230(08)\tabularnewline
\cline{2-4} 
 & 4 & $0.14\leq\varepsilon\leq0.3$ & 0.1881(18)\tabularnewline
\hline 
\multirow{4}{*}{0.4} & 1 & $0.16\leq\varepsilon\leq0.5$ & 0.3020(28)\tabularnewline
\cline{2-4} 
 & 2 & $0.16\leq\varepsilon\leq0.6$ & 0.3007(09)\tabularnewline
\cline{2-4} 
 & 3 & $0.16\leq\varepsilon\leq0.5$ & 0.2346(05)\tabularnewline
\cline{2-4} 
 & 4 & $0.16\leq\varepsilon\leq0.34$ & 0.1766(24)\tabularnewline
\hline 
\multirow{4}{*}{0.3} & 1 & $0.22\leq\varepsilon\leq0.6$ & 0.3216(26)\tabularnewline
\cline{2-4} 
 & 2 & $0.22\leq\varepsilon\leq0.7$ & 0.3150(06)\tabularnewline
\cline{2-4} 
 & 3 & $0.22\leq\varepsilon\leq0.5$ & 0.2230(08)\tabularnewline
\cline{2-4} 
 & 4 & $0.22\leq\varepsilon\leq0.4$ & 0.1578(21)\tabularnewline
\hline 
\multirow{4}{*}{0.2} & 1 & $0.26\leq\varepsilon\leq0.6$ & 0.3238(32)\tabularnewline
\cline{2-4} 
 & 2 & $0.26\leq\varepsilon\leq0.8$ & 0.3249(13)\tabularnewline
\cline{2-4} 
 & 3 & $0.26\leq\varepsilon\leq0.6$ & 0.2133(32)\tabularnewline
\cline{2-4} 
 & 4 & $0.26\leq\varepsilon\leq0.5$ & 0.1428(30)\tabularnewline
\hline 
\end{tabular}

\par\end{centering}

\protect
\caption{
The fitting range 
used in Fig.~\ref{fig:extrapolate_eps_3rd}
for the $\varepsilon \rightarrow 0$ extrapolations
is listed with the extrapolated values obtained by the fits.
\label{tab:The-list-m3}}
\end{table}

\end{document}